\documentstyle[epsfig]{mn}
\DeclareMathVersion{bold}
\newcommand{\erf}{\mathrm{erf}\mathnormal{}}

\title[Filter design]{Filter design for the detection of compact
sources based on the Neyman-Pearson detector} 

\author[L\'opez-Caniego et al.]{M. L\'opez-Caniego$^{1,2}$\footnotemark,
D. Herranz$^{3}$, R.B. Barreiro$^{1}$ and  J.L. Sanz$^{1}$ 
\\$^{1}$ Instituto de F\'\i sica de Cantabria, Avda. Los Castros s/n,
39905 Santander, Spain\\ 
$^{2}$ Depto. de F\'\i sica Moderna, Univ. de Cantabria, Avda. los
Castros, s/n/, 39005-Santander, Spain\\ 
$^{3}$ Istituto di Scienze e Tecnologie dell'Informazione
``A. Faedo'', CNR, via Moruzzi 1, 56124 Pisa, Italy} 
\date{Accepted 2005 February 24} 

\begin{document}

\maketitle

\begin{abstract}
This paper considers the problem of compact source detection on a Gaussian
background. We make a one-dimensional treatment (though a 
generalization to two or more dimensions is possible). 
Two relevant aspects of this problem are considered:
the design of the detector and the filtering of the data.
Our detection scheme is based on local maxima and it takes into
account not only the amplitude but also the curvature
of the maxima. A Neyman-Pearson
test is used to define the region of acceptance, 
that is given by a sufficient linear detector that is independent
on the amplitude distribution of the sources.
We study how detection can be enhanced by means of linear
filters with a scaling parameter
and compare some of them that have been proposed in
the literature (the Mexican Hat wavelet, 
the matched and the scale-adaptive filters). 
We introduce a new filter, that 
depends on two free parameters (biparametric
scale-adaptive filter). The value of these two parameters can be
determined, given the a priori {\it pdf} of the amplitudes of the sources,
such that the filter 
optimizes the performance of the detector
in the sense that it gives the maximum number 
of real detections once fixed the number density 
of spurious sources. 
The new filter includes as particular cases the standard matched
filter and the scale-adaptive filter.
Then, by construction, the biparametric scale adaptive filter
outperforms these filters.
The combination of a detection scheme that includes
information on the curvature and a flexible filter that 
incorporates two free parameters (one of them a scaling) 
improves significantly the number of detections in
some interesting cases. In particular, for the case of
weak sources embedded in white noise
the improvement with respect to the standard matched filter 
is of the order of $40\%$. Finally, an 
estimation of the amplitude of the source (most probable value) is introduced 
and it is proven that such an estimator is unbiased and it has 
maximum efficiency.
We perform numerical simulations to test
these theoretical ideas in a practical example and conclude that the
results of the simulations agree with the analytical ones.

\end{abstract}

\begin{keywords}
methods: analytical - methods: data analysis - techniques: image processing
\end{keywords}

\section{INTRODUCTION}
\footnotetext{E-mail: caniego@ifca.unican.es}

The detection of compact signals (sources) embedded in a background is 
a recurrent problem in many fields of Astronomy. Some common examples are
the separation of  
individual stars in a crowded optical image, the identification of
local features (lines) in noisy one-dimensional spectra
or the detection of faint
 extragalactic objects in microwave frequencies. 
The detection, identification and
removal of the extragalactic point sources (EPS) is fundamental for the
study of the Cosmic Microwave Background Radiation (CMB)  
data (Franceschini et al. 1989,
Toffolatti et al. 1998, de
Zotti et al. 1999). In particular, the contribution of EPS
is expected to be very relevant at the lowest and highest 
frequency channels of the future ESA Planck Mission (Mandolesi et al. 1998,
Puget et al. 1998).

The heterogeneous nature of the EPS that appear in CMB maps
as well as their unknown spatial distribution make difficult to 
separate them from the other physical components (CMB, Galactic dust,
synchrotron, etc)
by means of statistical component separation methods. 
Techniques based on the use of linear filters, however,
are well-suited
for the task of detecting compact spikes on a background.
Several techniques based on different linear filters have been proposed 
in the literature: the Mexican Hat Wavelet (MHW, Cay\'on et al. 2000, Vielva et
al. 2001a,b, 2003),
the classic \emph{matched} filter (MF, Tegmark and Oliveira-Costa 1998), the
Adaptive Top Hat
Filter (Chiang et al 2002) and the scale-adaptive filter (SAF, Sanz et al.
2001).
A certain deal of controversy has appeared about which one, if any, of the
previous
filters is \emph{optimal} for the detection of point sources in CMB data.

In order to answer that question it is necessary to consider first a more
fundamental
issue, the concept of \emph{detection} itself. 
The detection process can be posed as follows: given an observation, the
problem is to \emph{decide} whether or not a certain signal was present at the
input of the
receiver. The decision is not obvious since the observation is corrupted by 
a random process that we call `noise' or `background'.

 Formally, the \emph{decision}
is performed by choosing between two complementary hypotheses: that the 
observed data is originated by the background alone 
(\emph{null hypothesis}), and the hypothesis that
the observation corresponds to a combination of the background and the signal.
To decide, the detector should use all the available
information in terms of
the probabilities of both hypotheses given the data. The 
\emph{decision device} separates the space $\mathcal{R}$ of all possible 
observations in two disjoint subspaces, $\mathcal{R}_*$ and $\mathcal{R}_-$, 
so that if an observation 
$y \in \mathcal{R}_-$ the null hypothesis is accepted, and if $y \in
\mathcal{R}_*$ 
the null hypothesis is rejected, that is, a source is `detected'
($\mathcal{R}_*$ is called the region of acceptance).
Hence, we will call any generic decision device 
of this type a \emph{detector}.

The simplest example of detector, and one that has been
extensively used in Astronomy, is \emph{thresholding}: if the
intensity of the field is above a given value (e.g. 5$\sigma$), a
detection of the signal is accepted, on the contrary one assumes that
only background is present.
Thresholding has a number of advantages, among them the facts that it is 
straightforward and that
it has a precise meaning in the case of Gaussian backgrounds in the
sense of controlling the probability of spurious detections.
However, it does not use all the available 
information contained in the data to perform decisions. 
For example, the inclusion of spatial information (such as the
curvature) could help to distinguish 
the sources from fluctuations in the background with similar scale but a
different shape. 

A general detector that can use more information than simple
thresholding is given by the Neyman-Pearson (NP) decision rule:
\begin{equation}
L(x_i)=\frac{p(x_i|H_1)}{p(x_i|H_0)} \ge L_*
\end{equation}
where $L(x_i)$ is called the likelihood ratio, $p(x_i|H_0)$ is the
probability density function ({\it pdf}) associated to the null hypothesis
(i.e. there is no source) whereas $p(x_i|H_1)$ is the {\it pdf} corresponding to
the alternative hypothesis (i.e. there is a source). $x_i$ are a set
of variables which are measured from the data. $L_*$ is an arbitrary
constant, which defines the region of acceptance $\mathcal{R}_*$, and
must be fixed 
using some criterion. For instance, one can adopt a scheme for object
detection based on maxima. The procedure would consist 
on considering the intensity maxima of the image as candidates for
compact sources and apply to each of them the NP rule to decide
whether they are true or spurious.
For a 1D image, the ratio of probabilities would then correspond to
the probability of having a 
maximum with a given intensity and curvature (which are the variables
$x_i$ in this case) in the presence of background
plus signal over the probability of having a maximum when only
background is present. If this ratio is larger than a given value
$L_*$, the candidate is accepted as a detection, if not, it is rejected.

Unfortunately, in many cases the sources are very faint and this makes very
difficult to detect them. In order to improve the performance of the
detector, a prior processing of the image could be useful.
Here is where \emph{filtering} enters in scene. 
The role of filtering is to transform the data in such a way that
a detector can perform better than before filtering. 
Once the detector is fixed, it is interesting to compare the
performance of different filters, which has been rarely considered in
the literature.
In a recent work, Barreiro et al. (2003) introduce a novel technique
for the detection of sources based on the study of the number density
of maxima for the case of a Gaussian background in the presence or
absence of a source. In order to define the region of acceptance the
Neyman-Pearson decision rule is used with {\it pdf}'s associated to the
previous number densities and using the information of both the
intensities $\xi$ and the curvatures $\kappa$ of the peaks in a data
set. In addition, $L_*$ is fixed by maximising
the \emph{significance}, which is the weighted difference between the
probabilities of having and not having a source. In that work
the performances of several filters (SAF, MF and MHW) 
is compared in terms of their \emph{reliability},
defined as the ratio between the number density of true detections
over the number density of spurious detections. They find that,
on the basis of this quantity, the choice of the optimal filter
depends on the statistical properties of the background.

However, the criterion chosen to fix $L_*$ based on the significance
does not necessarily leads to the optimal reliability. 
Therefore, if we are considering the reliability as the main criterion
to compare filters, a different criterion for $L_*$,
based on number densities must be used.
In a posterior article, Vio et al. (2004), following the previous work,
adopt the same Neyman-Pearson decision rule, based on the {\it pdf}'s of
maxima of the background and 
background plus source, to define the region of acceptance. However,
they propose to find $L_*$ by fixing the number density of
spurious detections and compare the performance of the filters based
on the number density of true detections. In this case, the MF
outperforms the other two filters.
Note that in these last two works different criteria have been used to fix
$L_*$, thus leading to different results.

In the present work, our goal will also be to find an optimal filter
that gives a maximum number density of detections fixing a certain
number density of spurious sources. In order to define the
detector, we will use a decision rule based on the Neyman-Pearson
test. We will consider some standard filters (MF, SAF and MH) introduced in
the literature as well as a new filter that we call the Biparametric
Scale Adaptive Filter (BSAF). 
In all the filters appears in a natural way the scale of the
source. We will modify such a scale introducing  
an extra parameter. In fact, it has been shown by L\'opez-Caniego et
al. (2004) that the standard Matched Filter can be improved  
under certain conditions by filtering at a different scale from that
of the source. The performance of the BSAF will be compared with the other  
filters.

The overview of this paper is as follows.
In section 2, we introduce two useful quantities: number of maxima in a 
Gaussian background in the absence and presence of a local source. 
In section 3, we introduce the detection problem and define the region of
acceptance.  In section 4, we introduce an estimator of 
the amplitude of the source that is proven to be unbiased and maximum
efficient. In section 5 and 6, we obtain different analytical
and numerical results regarding weak point sources and scale-free
background spectra  
and compare the performance of the new filter with others used in the
literature. In section 7, we describe the simulations performed to test some 
theoretical aspects and give the main results and finally, 
in section 8, we summarize the conclusions and applications of this paper.
Appendix A is a sketch to obtain a sufficient linear detector whereas we 
obtain the linear unbiased and maximum efficient estimator in appendix B.

\section{BACKGROUND PEAKS AND COMPACT SOURCES}

\subsection{The background}

\begin{figure} 
\epsfxsize=84mm
\epsffile{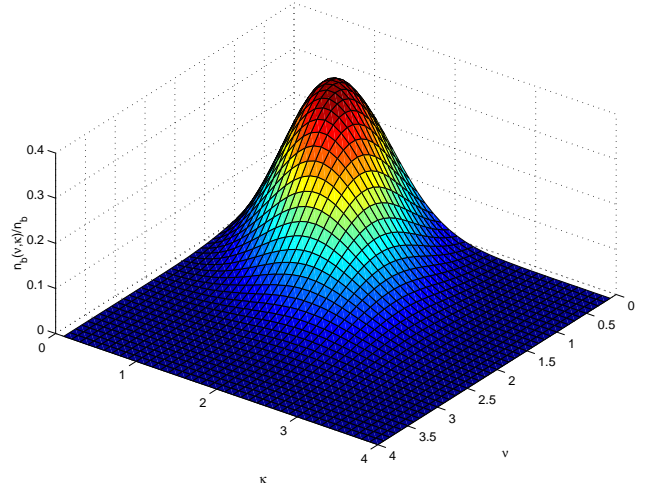}
\caption{Value of  $n_b(\nu ,\kappa )/n_b$ for $\rho=0.7$.}
\label{fig:fig1}
\end{figure}

Let us assume a 1D background (e. g. one-dimensional scan on the
celestial sphere or time 
ordered data set) represented by a Gaussian random field $\xi (x)$
with average value  
$\langle \xi (x)\rangle = 0$ and power spectrum $P(q), \ q\equiv |Q|$: 
$\langle \xi (Q)\xi^* (Q')\rangle = P(q)\delta_D (Q - Q')$, where
$\xi (Q)$ is the Fourier transform of $\xi (x)$ and $\delta_D$ is the 1D Dirac 
distribution. 
The distribution of
maxima was studied by Rice (1954) in a pioneering article. The expected
number density of 
maxima per intervals $(x, x + dx)$, $(\nu ,\nu + d\nu )$ and $(\kappa
,\kappa + d\kappa )$ is given by
\begin{equation} 
n_b(\nu ,\kappa )  = \frac{n_b\,\kappa}{\sqrt{2\pi (1-\rho^2)}} 
e^{- \frac{\nu^2 + \kappa^2 - 2\rho \nu \kappa}{2(1 - \rho^2)}}, 
\label{nbackground}
\end{equation}
being $n_b$ the expected total number density of maxima (i.e. number of 
maxima per unit interval $dx$)
\begin{equation}  
n_b \equiv \frac{1}{2\pi \theta_m},\ \ \ 
\nu \equiv \frac{\xi}{\sigma_0},\ \ \ 
\kappa \equiv \frac{-\xi^{\prime \prime}}{\sigma_2}, \\ 
\end{equation}

\begin{equation}
\theta_m \equiv \frac{\sigma_1}{\sigma_2},\ \ \ 
\rho \equiv \frac {\sigma_1^2}{\sigma_0 \sigma_2} = \frac{\theta_m}{\theta_c},\
\ \ 
\theta_c \equiv \frac{\sigma_0}{\sigma_1}, \nonumber 
\end{equation}
where $\nu \in (-\infty ,\infty )$ and $\kappa \in [0,\infty )$ represent the 
normalized field and curvature, 
respectively. $\sigma_n^2$ is the moment of order
$2n$ associated to the field. $\theta_c ,\theta_m $ are the coherence scale of
the field
and maxima, respectively. 
As an example, figure~\ref{fig:fig1} shows 
the values of the ratio $n_b(\nu ,\kappa )/n_b$ for the case $\rho=0.7$ (a typical value for the backgrounds we are considering). 
In this case, the expected density of maxima has a peak 
around $\nu \simeq 0.8$ and $\kappa \simeq 1.1$, that is, most of the peaks 
appear at a relatively low threshold and curvature, and the density of peaks
decreases quickly for extreme values of $\nu$ and $\kappa$.

If the original field is linear-filtered with a circularly-symmetric 
filter $\Psi(x;R,b)$, dependent on $2$ parameters ($R$ defines a scaling whereas $b$ defines a translation)

\begin{equation}
\Psi (x; R, b) = \frac{1}{R}\psi \left(\frac{|x - b|}{R}\right),
\end{equation}
we define the filtered field as
\begin{equation}
w(R, b) = \int dx\,\xi (x)\Psi (x; R, b).
\end{equation}
Then, the moment of order $2n$ of the linearly-filtered field is
\begin{equation}
\sigma_n^2 \equiv 2\int_0^{\infty} dq\,q^{2n}P(q)\psi^2 (Rq),
\end{equation}
being $P(q)$ the power spectrum of the unfiltered field and $\psi (Rq)$ 
the Fourier transform of the circularly-symmetric linear filter.

\subsection{The presence of a local source}

Now, let us consider a position $x$ in the image where a Gaussian source
(i.e. profile given by  
$\tau (x) = \exp ({- x^2/2R^2})$, where $R$ is the beam width) is
superimposed to the previous background.
Then, the expected number density of maxima per intervals $(x, x + dx)$,
$(\nu ,\nu + d\nu )$ and  
$(\kappa ,\kappa + d\kappa )$, given a source of amplitude $A$ in such
spatial interval, is given by (Barreiro et al. 2003)
\begin{eqnarray}\label{nsource}
n(\nu ,\kappa |\nu_s) =  \frac{n_b\,\kappa}{\sqrt{2\pi (1 - \rho^2)}} \times \nonumber
\end{eqnarray}
\begin{eqnarray}
\indent e^{- \frac{(\nu - \nu_s)^2 + (\kappa -\kappa_s)^2 -  2\rho (\nu - \nu_s)(\kappa - \kappa_s)}{2(1 - \rho^2)}},
\end{eqnarray}
where $\nu \in (-\infty ,\infty )$ and $\kappa \in [0,\infty )$, 
$\nu_s = A/\sigma_0$ is the normalized amplitude of the source and 
$\kappa_s = - A\tau_{\psi}^{\prime \prime}/\sigma_2$ is the
normalized curvature of the filtered source.
The last expression can be obtained as
\begin{equation}
\kappa_s = \nu_s y_s,~ 
y_s \equiv - \frac{\theta^2_m \tau_{\psi}^{\prime \prime}}{\rho},~
- \tau_{\psi}^{\prime \prime} = 2\int_0^{\infty}dq\,q^2\tau (q)\psi (Rq).
\end{equation}
Note that due to the statistical homogeneity and isotropy of the background, the
previous equations are independent of the position of the source.

We consider that the filter is normalized such that the amplitude of
the source is the same after linear filtering: $\int dx\,\tau (x)\Psi
(x; R, b) = 1$. 

\section{THE DETECTION PROBLEM}

We want to choose between different filters based on \emph{detection}.
To make such a decision, we will focus on the following two 
fundamental quantities:
a) the number of spurious sources which emerge after the filtering 
and detection processes 
and b) the number of real sources detected. 
As we will see in this section, these quantities are
properties of the Gaussian field and source that can be calculated
from equations (\ref{nbackground}) and (\ref{nsource}).
As we will see, the previous 
properties are not only related to the SNR gained in the filtering 
process but depend on the filtered momenta up to 4th-order (in the 1D case), i.e. 
the amplification and the normalized curvature.

Let us consider a local peak in the 1D data set characterised by the 
normalized amplitude and curvature $(\nu_s ,\kappa_s)$.
Let
$H_0$: n.d.f. $n_b(\nu ,\kappa )$ represents the \emph{null}
hypothesis, i.e. the local number density of background maxima, and 
$H_1$: n.d.f.  $n(\nu ,\kappa )$ represents the \emph{alternative}
hypothesis, i.e. the local number density of maxima when there is a compact
source: 
\begin{equation}
n (\nu ,\kappa )= \int_0^{\infty} d\nu_s \, p(\nu_s)n(\nu ,\kappa |\nu_s ).
\label{eq:number_b+s}
\end{equation}
In the previous equation, we have introduced a priori
information about the probability distribution of the sources:
we get the number density of source detections weighting with the a
priori probability $p(\nu_s)$.

To construct our detector, we will assume a Neyman-Pearson (NP)
decision rule using number densities instead of probabilities:
\begin{equation}
L(\nu ,\kappa )\equiv \frac{n(\nu ,\kappa )}{n_b(\nu ,\kappa )}\geq L_*,
\label{lik_nus}
\end{equation}
where $L_*$ is a constant.
The previous equation defines a region in $(\nu ,\kappa)$, the so-called
region of acceptance $\mathcal{R}_*$. 
Therefore, the decision rule is expressed such that if the values of
$(\nu ,\kappa)$ of the candidate maximum is inside $\mathcal{R}_*$ 
(i.e. $L\geq L_*$) we decide that the signal is 
present. On the contrary, if $L< L_*$ we decide that the signal is absent.
\footnote{Note that the region defined by equation (\ref{lik_nus}) is
equivalent to the one defined by the 
usual Neyman-Pearson test in terms of probabilities 
\begin{equation}
\frac{p(\nu,\kappa)}{p_b(\nu,\kappa)} \ge L_*^\prime
\label{eq:np}
\end{equation}
where $p_b(\nu,\kappa)$, $p(\nu,\kappa)$ are
the {\it pdf}'s associated to the number densities given by equations 
(\ref{nbackground}) and (\ref{eq:number_b+s}) and, in order to compare
different filters, the constant $L_*^\prime$ must be found by
fixing the number density of spurious sources in the region of
acceptance instead of the {\it false alarm} probability.} 

It can be proved that the previous region of
acceptance $\mathcal{R}_*$ is equivalent to the sufficient linear
detector (see Appendix A)
\begin{equation}
\mathcal{R}_*:  \varphi (\nu ,\kappa )\geq \varphi_* ,
\label{eq:r_*}
\end{equation}
where $\varphi_*$ is a constant and $\varphi$ is given by 
\begin{equation} 
\varphi (\nu ,\kappa )\equiv \frac{1 - \rho y_s}{1 - \rho^2}\nu +
\frac{y_s - \rho }{1 - \rho^2} \kappa
\label{eq:phi}
\end{equation}
We remark that the assumed criterion for detection leads to a
\emph{linear} detector $\varphi$ (i.e. linear dependence on the
threshold $\nu$ and curvature $\kappa$). Moreover, this detector is
independent of the {\it pdf} of the source amplitudes.

Using this detector, the expected number density of spurious sources and of
true detections are given by
\begin{equation} \label{eq:nbstar}
n_b^* = \int_{\mathcal{R}_*}d\nu \,d\kappa \,n_b(\nu ,\kappa), 
\end{equation}
\begin{equation} \label{eq:nstar}
n^* = \int_{\mathcal{R}_*}d\nu \,d\kappa \,n(\nu ,\kappa).
\label{eq:ndet}
\end{equation}

We remark that in order to get the true number of real source
detections such a number must be multiplied by the probability to have
a source in a pixel in the data set.

Note that for a fixed number density of spurious sources $n_b^*$, the
NP detector leads to the maximum number density of true detections
$n^*$. 

Taking into account equations (\ref{eq:r_*}) to (\ref{eq:ndet}), one
can find $n_b^*$ and $n^*$ for a Gaussian background. After a
straightforward calculation, the number density of spurious sources
found using the NP rule is given by:
\begin{eqnarray} 
n_b^* &=& \frac{n_b}{2}\left[ {\rm erfc}\left(\frac{\varphi_*\sqrt{1-\rho^2}}
{\sqrt{2}(1-\rho y_s)}\right) \right. \nonumber \\
&& \left. +\sqrt{2}M y_s {\rm e}^{-M^2
\varphi_*^2} {\rm erfc}\left( -\frac{\sqrt{1-\rho^2}}{1-\rho y_s} y_s
M \varphi_* \right)\right],
\label{eq:nb*}
\end{eqnarray}
\begin{eqnarray}
M \equiv \sqrt{\frac{1-\rho^2}{2(1-2\rho y_s+y_s^2)}}. \nonumber
\end{eqnarray}

Similarly, the number density of detections is obtained as:
\begin{eqnarray}  
n^*  & = & \frac{n_b}{\sqrt{2\pi}}\frac{1 - \rho y_s}{(\mu +
y_s^2)\sqrt{1 - \rho^2}} \times \nonumber \\
 & & \int_{\varphi_*}^\infty d\varphi \, I(\varphi )[1 + B(z)]e^{-
\frac{(1 - \rho^2)\varphi^2}{2{(1 - \rho y_s)}^2}}, 
\label{eq:nb}
\end{eqnarray}

\noindent where
\begin{eqnarray}
z = \frac{y_s\varphi}{1 - \rho y_s}\sqrt{\frac{1 - \rho^2}{2(\mu +
y^2_s)}},\nonumber  
\end{eqnarray}

\begin{eqnarray}
B(z)=\sqrt{\pi} z e^{z^2} {\rm erfc}(-z), \, \, \, \,
\mu \equiv \frac{{(1 - \rho y_s)}^2 }{1 - \rho^2}, \nonumber
\end{eqnarray}
\begin{eqnarray}
I(\varphi ) = \int_0^{\infty} d\nu_s\,p(\nu_s)e^{\nu_s\varphi-\frac{1}{2}\nu^2_s(\mu + y^2_s)}.  
\label{nstar2}
\end{eqnarray}

\section{The estimation of the amplitude of the source}

The signal has an unknown parameter, the amplitude $A$, that has to 
be estimated from the data $(\nu ,\kappa )$. We shall 
assume that the most probable value of the distribution
$n(\nu ,\kappa |\nu_s )$ gives an estimation of the amplitude of the source
(criterion for amplitude estimation). 
The result $\hat{\nu}_s$ is given by the equation
\begin{equation} \label{eq:estim_amplitude}
\hat{\nu}_s =  \frac{\varphi (\nu ,\kappa )}{y^2_s + \mu},
\end{equation}    
where the function $\varphi$ is given by equation (\ref{eq:phi}).
One can prove that the previous expression corresponds to a linear
estimator that is unbiased and maximum efficient (minimum variance),
i.e.
\begin{equation}
\langle {\hat{\nu}}_s \rangle = \nu ,\ \ \ \sigma^2_{{\hat{\nu}}_s} =
\frac{1}{y^2_s + \mu},
\end{equation}
where $\langle \rangle$ denotes average value over realizations (see
Appendix B). 

\section{ANALYTICAL RESULTS}

\subsection{Filters}

We will consider as an application the detection of compact sources characterised by a 
Gaussian profile $\tau (x) = \exp (- x^2/2R^2)$, and Fourier transform
$\tau = R \exp (-(qR)^2/2)$, though the extension to other  
profiles will be considered in the future. Such a profile is physically and astronomically
interesting because it represents the convolution of a point source (Dirac $\delta$ distribution) with
a Gaussian beam. 

The source profile above includes a ``natural scale'' $R$ that
characterises the source. This is a fundamental scale  
that will appear in all the filters we will consider here. 
By construction, the standard MF and SAF operate on  
this scale, as well as the canonical MHW at the scale of the source. However, 
it has been shown that changing the scale at which the MHW and the MF
filter the image can improve its performance in terms of  
detection (Vielva et al. 2001a, L\'opez-Caniego et
al. 2004). Following this idea, 
we will introduce another degree of freedom in all the filters
that allows us to change their scale in a  
continuous way (similarly to the scaling of a continuous
wavelet). This degree of freedom is obtained by multiplying  
the scale $R$ by a new parameter $\alpha > 0$. We will show that with
this new parameter the improvement in the results is significant.

 \subsubsection{The scale-adaptive filter (SAF)}

\begin{figure} 
\epsfxsize=85mm
\epsffile{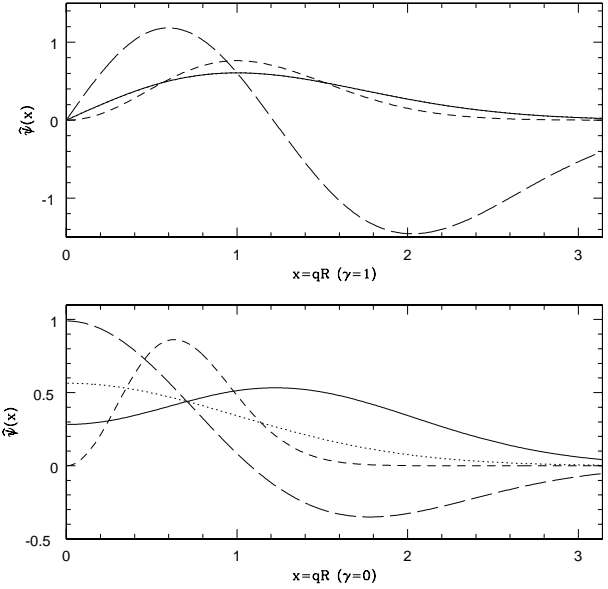}
\caption{Different filters for the values of $\gamma=0$ (lower panel) and 
$\gamma=1$ (upper panel). The filters represented in all cases are: the SAF (solid line),
MF (dotted line),  MH (short dashed line), which are shown for
$\alpha$ = 1, and the BSAF (long dashed line) which is given  
for $(\alpha ,c) = (0.3,-0.86)$ for $\gamma=0$ and  $(\alpha ,c) = (0.4,-0.68)$ for $\gamma=1$. Note 
that for $\gamma = 1$ the SAF and the MF coincide.}
\label{fig:fig3}
\end{figure}

The idea of a scale-adaptive filter (or optimal pseudo-filter) has been
recently introduced by the authors (Sanz et al. 2001).
By introducing a circularly-symmetric filter, $\Psi (x; R, b)$, 
we are going to express the conditions in order to obtain a
scale-adaptive filter for the 
detection of the source $s(x)$ at the origin taking into account the
fact that the source is 
characterised by a single scale $R_o$. The following conditions are assumed: 
$(1) \  \langle w(R_o, 0)\rangle = s(0) \equiv A$, i.e. $w(R_o, 0)$ is an
\emph{unbiased} 
estimator of the amplitude of the source;
$(2)$ the variance of $w(R, b)$ has a minimum at the scale $R_o$,
i.e. it is an \emph{efficient} estimator;
$(3)$  $w(R, 0)$ has a maximum with respect to the scale at $(R_o, 0)$.
Then, the filter satisfying these conditions is given by (Sanz et al. 2001)

\begin{eqnarray} \label{eq:saf}
\psi_{SAF} = \frac{1}{ac - b^2} \frac{\tau (q)}{P(q)}
\left[b + c - (a + b)\frac{dln\tau}{dlnq}\right], \nonumber
\end{eqnarray}
\begin{eqnarray}
a\equiv \int dq\,\frac{{\tau}^2}{P},\ 
b\equiv \int dq\,q \frac{\tau}{P}\frac{d\tau}{dq},\  
c\equiv \int dq\,q^2\frac{1}{P}{\left(\frac{d\tau}{dq}\right)}^2,
\end{eqnarray}

Assuming a scale-free power spectrum, $P(q)\propto q^{- \gamma}$, and a
Gaussian profile for the source, the previous set of equations lead to
the filter 
\begin{eqnarray}\label{eq:saf_m}
\tilde{\psi}_{SAF} = N(\alpha)x^{\gamma}e^{- \frac{1}{2}x^2}\left[1+ \frac{t}{m^2}x^2\right], \ \ x\equiv q\alpha R, \nonumber
\end{eqnarray}
\begin{eqnarray}
m\equiv \frac{1 + \gamma}{2}, \ \  t\equiv \frac{1 - \gamma}{2}, \ \ \Delta \equiv \frac{2\alpha^2}{1+\alpha^2},\nonumber
\end{eqnarray}
\begin{eqnarray}
N(\alpha) = \frac{\alpha}{\Delta^m \Gamma(m)}\frac{1}{1+\frac{t}{m}\Delta},
\end{eqnarray}
\noindent where we have modified the scale as $\alpha R$. 

In this case the filter parameters $\theta_m , \ \rho$ and the
curvature of the source $y_s$ are given by
\begin{eqnarray}
\frac{\theta_m}{\alpha R} =  \sqrt{\frac{1 + \frac{t^2}{m} + \frac{2t}{m^2}}
{\left(1 + m\right)\left(1 + \frac{t^2}{m} + \frac{2t(2 + t)}{m^2}\right)}} \nonumber
\end{eqnarray}
\begin{eqnarray}
\rho = \sqrt{\frac{m}{1+m}}
\frac{1 + \frac{t^2}{m} + \frac{2t}{m^2}}{\sqrt{\left(1 + \frac{t^2}{m}\right)\left(1 + \frac{t^2}{m} + \frac{2t(2 + t)}{m^2}\right)}},\nonumber  
\end{eqnarray}
\begin{eqnarray}
y_s = H \sqrt{\frac{1 + \frac{t^2}{m}}{m\left(1 + m\right)\left(1 + \frac{t^2}{m} + \frac{2t(2 + t)}{m^2}\right)}},\nonumber
\end{eqnarray}
\begin{eqnarray}
H \equiv \Delta \frac{1+\frac{t}{m}}{1 + \frac{t}{m}\Delta} \frac{m^2 + t(1+m)\Delta}{m^2 + t(1+m)}
\end{eqnarray}

Figure~\ref{fig:fig3} shows the SAF for two values of the spectral index.

\subsubsection{The matched filter (MF)}

If one removes condition (3) defining the SAF
in the previous subsection, it is not difficult to find another type of
filter after minimization of the variance (condition  
(2)) with the constraint (1)
\begin{equation} \label{eq:mf}
\psi_{MF} = \frac{1}{2a}\frac{\tau (q)}{P(q)}.
\end{equation} 
\noindent This will be called \emph{matched} filter as is usual in
the literature. Note that in general the matched and adaptive filters
are different.

For the case of a Gaussian profile for the source and a scale-free power 
spectrum given by $P(q)\propto q^{-\gamma}$, the previous formula leads to the 
following modified matched filter
\begin{equation} 
\tilde{\psi}_{MF} = N(\alpha)x^{\gamma}e^{- \frac{1}{2}x^2},\ \ \
x\equiv q\alpha R, 
\end{equation}
\begin{displaymath}
N(\alpha)= \frac{\alpha}{\Delta^m \Gamma(m)}
\end{displaymath}
where $m$ and $\Delta$ is given by equation (\ref{eq:saf_m}) and we
have included the scale parameter $\alpha$.

Figure~\ref{fig:fig3} shows the MF for the case $\alpha=1$ (standard MF) and values of the
spectral index $\gamma = 0, 1$. We remark that for $\gamma=1$ the scale-adaptive
filter and the matched filter coincide, and for $\gamma=2$ (not shown in the figure), the matched filter and the Mexican Hat wavelet are equal.

For the MF the parameters $\theta_c ,  \ \theta_m , \ \rho$ and the curvature of
the
source $y_s$ are given by
\begin{equation}
\frac{\theta_m}{\alpha R} =  \frac{1}{\sqrt{1 + m}}  ,\ \ \ 
\rho = \sqrt{\frac{m}{1 + m}},\ \ \ 
y_s = \rho \Delta
\end{equation}
We remark that the linear detector $\varphi (\nu ,\kappa )$ is reduced to $\varphi=\nu $ for the standard Matched Filter ($\alpha =1$).
i.e. curvature does not affect the region of acceptance for such a filter.

\subsubsection{The mexican hat wavelet (MH)}

The MH is defined to be proportional to the Laplacian of the
Gaussian function:
\begin{equation} 
\psi_{MH} (y) \propto (1 - y^2) e^{- y^2/2}. 
\end{equation}
Thus, in Fourier space
\begin{equation}              
\psi_{MH} (q) = \frac{2}{\sqrt{\pi}} {(qR)}^2e^{- \frac{1}{2}{(qR)}^2}.
\end{equation}

In this case the filter parameters $\theta_m , \ \theta_c , \ \rho$ and the
curvature of the source $y_s$ are given by
\begin{equation}
\frac{\theta_m}{R} =  \frac{1}{\sqrt{3 + t}}, \ \ 
\rho = \sqrt{\frac{2+t}{3+t}}, \ \ 
y_s =  \frac{3/2}{\sqrt{(2+t)(3+t)}}.
\end{equation}
The generalization of this type of wavelet
for two dimensions has been extensively used for point source detection in 2D
images (Cay\'on et al. 2000, Vielva et al. 2001, 2003). As for the
previous filter, the MH is modified by including the scale parameter
$\alpha$ in the form 
\begin{equation}
\tilde{\psi}_{MH}  = N(\alpha)x^2e^{- \frac{1}{2}x^2},\ \ \ x\equiv q\alpha R, \ \
\end{equation}
\begin{displaymath}
N(\alpha)= \frac{2 \alpha}{\sqrt{\pi} \Delta^{3/2}}
\end{displaymath}

For the MH the parameters $\theta_c , \ \theta_m , \ \rho$ and the curvature of
the source $y_s$ are given by
\begin{equation}
\frac{\theta_m}{\alpha R} =  \frac{1}{\sqrt{3 + t}}, \ \ \rho = \sqrt{\frac{2+t}{3+t}},\ \  
y_s = \frac{3\Delta /2}{\sqrt{(2+t)(3+t)}}.
\end{equation}

Figure~\ref{fig:fig3} shows  the MH for different values of the spectral index.

\subsubsection{The biparametric scale adaptive filter (BSAF)}

If one removes condition (3) defining the SAF in subsection 5.1.1 and
introduces the condition that $w(R_o,b)$ has  
a spatial maximum in the filtered image at $b=0$, i.e. $\langle
w^{\prime \prime}
(R_o,0) \rangle$  $<0$, it is not difficult to find another type of
filter 
\begin{equation}
\psi \propto \frac{\tau(q)}{P(q)}(1+c(qR)^2),
\label{eq:eqnndf1}
\end{equation}
\noindent where $c$ is an arbitrary constant that can be related to the curvature of the maximum. We remark that the 
constraint $\langle w^\prime (R_o,0) \rangle = 0$ is automatically satisfied for any circularly-symmetric filter if the 
source profile has a maximum at the origin.

For the case of a scale-free power spectrum, the filter is given by the parametrized equation
\begin{equation}
\tilde{\psi}_{BSAF} = \frac{\alpha}{2J_\gamma}x^{\gamma}e^{- \frac{1}{2}x^2}(1+cx^2),\ \ x\equiv q\alpha R.
\label{eq:eqnndf}
\end{equation}

\noindent where we have modified the scale as $\alpha R$. Hereinafter, we will call this new filter containing two arbitrary parameters, $\alpha >0$ and 
$c$, the biparametric scale-adaptive filter (BSAF).

A calculation of the different moments leads to
\begin{eqnarray}  
\frac{\theta_m}{\alpha R} = \sqrt{\frac{G_{\gamma + 2}}{G_{\gamma + 4}}},  \
\rho = \frac{G_{\gamma + 2}}{\sqrt{G_{\gamma}G_{\gamma + 4}}}, \   
y_s = \frac{J_{\gamma + 2}}{J_{\gamma}}\sqrt{\frac{G_{\gamma}}{G_{\gamma + 4}}},
\end{eqnarray}
where m and t are defined in equation (\ref{eq:saf_m}) and $G_{\gamma}$ and $J_{\gamma}$ are given by

\begin{equation}
G_{\gamma} \equiv \frac{1}{2}[1 + 2mc + m(m+1)c^2]\Gamma (m),
\end{equation}

\begin{equation}
J_{\gamma}(\alpha) \equiv \frac{1}{2}[1 + mc\Delta] \Delta^m \Gamma (m).
\end{equation}

Note that the BSAF contains all the other considered filters as particular cases: 
the MF is recovered for $c=0$, when $c= t/m^2$ the BSAF defaults to 
the SAF and, finally, the MH wavelet is obtained in the two cases: 
$\gamma =0$,  $c \gg 1$ and  $\gamma =2$, $c = 0$.

\subsection{A priori probability distributions $p(\nu_s )$}

We will test two different {\it pdf} $p(\nu_s )$: a uniform distribution 
in the interval $0 \leq \nu \leq \nu_c$ and a scale-free distribution
with a lower and upper cut-off $\nu_i \leq \nu \leq \nu_f$. 
In particular, we will especially focus on values for the cut-off's
that lead to distributions dominated by weak sources. It is in this
regime where sophisticated detection methods are needed, since bright
sources can be easily detected with simple techniques.

\subsubsection{Uniform distribution}

In this case, 
\begin{equation}
p(\nu_s ) = \frac{1}{\nu_c}, \ \ \ \nu_s\in [0, \nu_c]. 
\end{equation}
This allows us to obtain
\begin{displaymath}
I(\varphi ) = \sqrt{\frac{\pi}{2}}\frac{e^{h^2}}{\nu_c\sqrt{y^2_s + \mu }}
\left[ \erf(h) + \erf \left (\frac{\nu_c}{\sqrt{2}}\sqrt{y^2_s + \mu} -h \right)
\right],
\end{displaymath}
\begin{equation}
h\equiv \frac{\varphi }{\sqrt{2(y^2_s + \mu )}}.
\end{equation}

In general, we will consider a cut-off in the amplitude of the sources
such that $\nu_c=2$ after filtering with he standard MF. Note 
that this correspond to different thresholds for the rest of the filters.

\subsubsection{Scale-free distribution with lower and upper cut-off}

In this case,
\begin{equation}
p(\nu_s ) = N\nu_s^{-\beta}, \ \ \ \nu\in [\nu_i, \nu_f ],\ \ \beta \neq 1,
\end{equation}
where the normalization constant N and $I(\varphi)$ are 

\begin{equation} 
N = \frac{\beta - 1}{\nu_i^{1-\beta}}\frac{1}{1-\left(\frac{\nu_i}{\nu_f} \right)^{\beta-1}}
\end{equation}

\begin{equation}
I(\varphi ) = N\int_{\nu_i}^{\nu_f}d\nu\,\nu^{-\beta}e^{\nu[\varphi - \frac{\nu}{2}(\mu+y_s)^2]}.\\ 
\end{equation}

In general, we will consider $\beta=0.5$ and $\nu_i= 0.5$,
$\nu_f= 3$ after filtering with the standard MF and the
corresponding thresholds for the other filters.

\section{NUMERICAL RESULTS}

For a fixed number density of spurious sources $n^*_b$, we want to find the
optimal filter that produces the maximum number density of true
detections $n^*$ for different spectral indices ($\gamma$), values of
$R$ and point source distributions. 
In order to do this, we first obtain implicitly the value of
$\varphi_*$ from equation 
(\ref{eq:nb*}) (for a fixed value of  $n^*_b$) and then substitute it
in equation (\ref{eq:nb}) to calculate $n^*$. 
We consider two different distributions of sources to test
the robustness of the method: a  
uniform distribution and a scale-free distribution. 
Given that bright point sources are relatively easy to detect, we
mainly concentrate on the more interesting case of weak sources. In
any case, we also mention some results for distributions
containing bright sources.

We remark that the BSAF has an additional degree of freedom, the
parameter $c$, as it appears in equation (\ref{eq:eqnndf}). Note that
the BSAF and the SAF are not the same filter. The parameter $c$ in the
BSAF can take any positive or negative value,  
while the coefficient $t/m^2$, for the SAF, is a known function of
$\gamma$. By construction, the BSAF always outperforms
the MF and SAF or, in the worst case, defaults to the
best of them. 

\subsection{Uniform distribution}

\subsubsection{Weak sources}

\begin{figure}
\epsfxsize=75mm
\epsffile{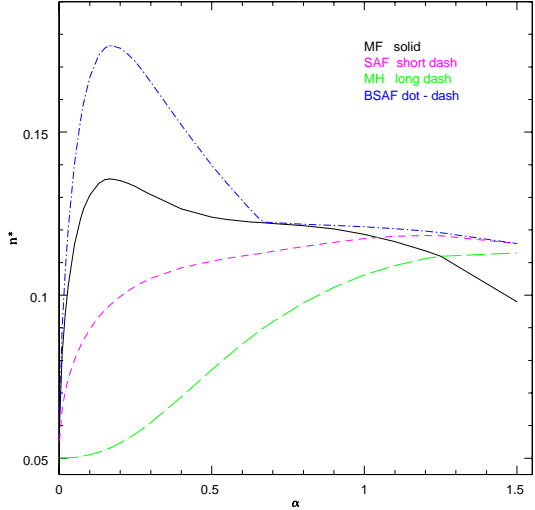}
\caption{Uniform  distribution. The expected number density of
detections $n^*$ as a function of the filter parameter $\alpha$ for  
$\gamma=0$  for the BSAF (where c has been obtained by maximising the
number of detections for each value of $\alpha$),  
MF, SAF and MH filters. We consider the case $R=3$, $n^*_b=0.05$ and
$\nu_c=2$.} 
\label{fig:numdet_alfa_unif_g0}
\end{figure}
\begin{figure}
\epsfxsize=75mm
\epsffile{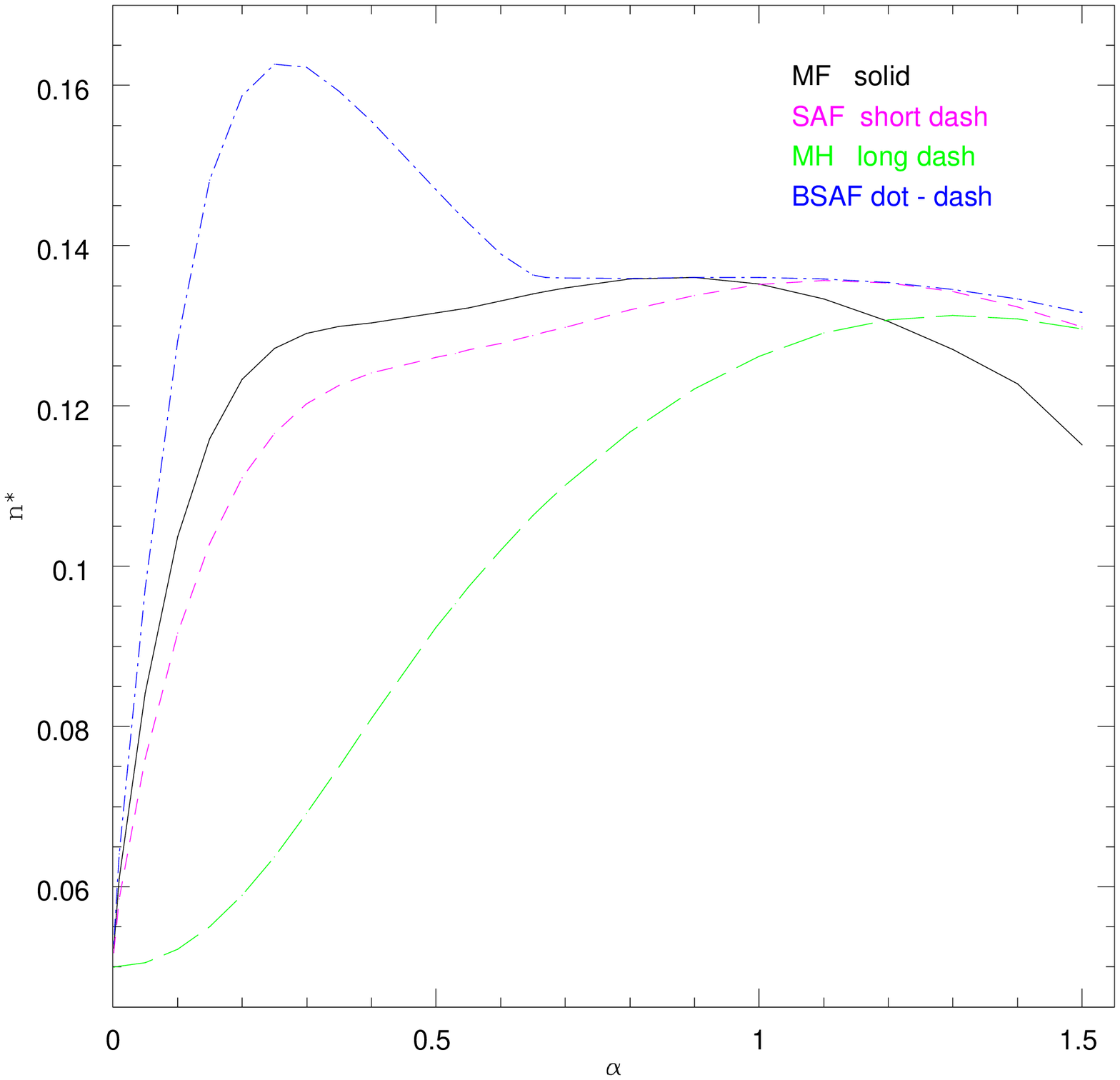}
\caption{Uniform  distribution. The expected number density of
detections $n^*$ as a function of the filter parameter $\alpha$ for  
$\gamma=0.5$  for the BSAF, MF, SAF and MH filters. We consider the
case $R=3$, $n^*_b=0.05$ and $\nu_c=2$. As in the previous figure  
the parameter c of the BSAF has been determined by maximising the
number of detections for each value of $\alpha$.}  
\label{fig:numdet_alfa_unif_g05}
\end{figure}
\begin{figure}
\epsfxsize=75mm
\epsffile{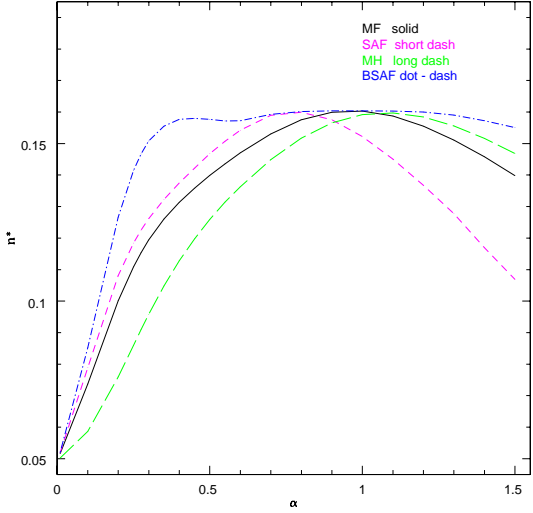}
\caption{Uniform  distribution. The expected number density of
detections $n^*$ as a function of the filter parameter $\alpha$ for  
$\gamma=1.5$  for the BSAF, MF, SAF and MH filters. We consider the
case $R=3$, $n^*_b=0.05$ and $\nu_c=2$. As in the previous figure  
the parameter c of the BSAF has been determined by maximising the
number of detections for each value of $\alpha$.} 
\label{fig:unif_g1p5}
\end{figure}

As a first case, we consider a uniform distribution of sources with
amplitudes in the interval $A\in[0,2]\sigma_0$, 
where $\sigma_0$ is the zero-order moment of the linearly-filtered map
with the standard MF. Therefore, the threshold $\nu$ in the image filtered
with this filter is in the interval $[0,2]$. Thus, 
the corresponding upper limit for $\nu$ in the original (unfiltered) map is
below 2, what means that we are considering the detection of weak sources.

As a reference example, in figure~\ref{fig:numdet_alfa_unif_g0}, we plot
$n^*$, the number 
density of detections, as a function of $\alpha$ for the case
$\gamma=0$, $n^*_b=0.05$ and $R=3$, where $R$ is given in pixel units.
For completeness, the theoretical values of $n^*$ are given, in
this figure, for values of $\alpha$ down to zero (note that $n^*$ $\to$
$n^*_b$ when $\alpha \to 0$). However, from a practical point of view, 
we do not expect the theoretical results to reproduce
the values obtained for a pixelized image when filtering at small
scales (since the effect of the
pixel is not taken into account). Therefore, hereinafter, we will only
consider those results obtained when filtering at scales larger (or of
the order) of the pixel size, which corresponds to $\alpha \ga
R^{-1}$. Taking into account this constraint, the best results are
obtained for $\alpha \simeq 0.3$ for the BSAF, that clearly
outperforms the standard MF (i.e., $\alpha=1$) with an improvement of
the $40\%$ in $n^*$. If we compare with the MF at
$\alpha=0.3$, the improvement is of $\simeq 20\%$. 

In figure~\ref{fig:numdet_alfa_unif_g05}, we give the same results for
the case $\gamma=0.5$. In this case, the BSAF at $\alpha=0.3$
improves again significantly the standard MF, with an
increase in the number density of detections of $\sim 25\%$.

As $\gamma$ increases, the improvement of the BSAF with respect to the
standard MF decreases. In fact, for values of $1 < \gamma \le 2$
they produce very similar results. As an example, we give the number
of detections achieved for each filter for the case $\gamma=1.5$,
$n_b^*=0.05$ and $R=3$ in Fig.~\ref{fig:unif_g1p5}. It can be seen
that the maximum number of detections is approximately found for the
standard MF. However, we would like to point out that the SAF and MH
wavelet at the optimal scale give approximately the same number of
detections as the standard MF. These results  
show the importance of filtering at scales $\alpha R$ instead of the usual 
scale of the source $R$.

\begin{figure} 
\epsfxsize=75mm
\epsffile{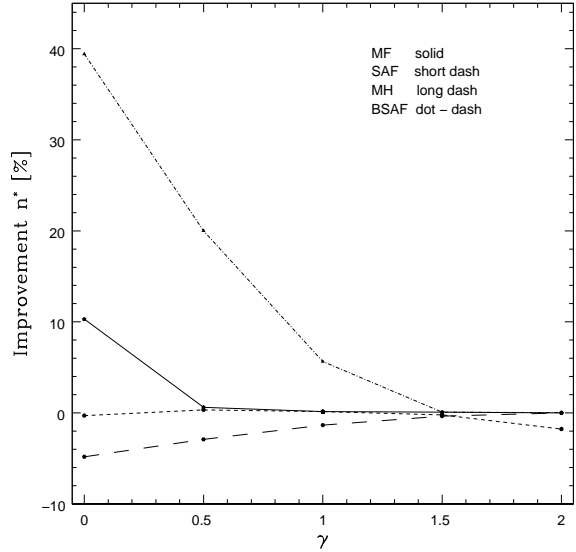}
\caption{Relative difference in the number of detections, 
with respect to the standard MF, as a function of the spectral
index $\gamma$. Values of $n_b^*=0.05$ and $R=3$ are used for a
uniform distribution of weak sources ($\nu_c=2$). For each filter,
the results are shown for the optimal parameters. 
}
\label{fig:unif_gamma}
\end{figure}

This can also be seen in Fig.~\ref{fig:unif_gamma}, that summarizes
how the relative 
performance of the considered filters with respect to the standard MF
changes with the spectral index $\gamma$ (again for $n_b^*=0.05$ and
$R=3$). For each filter, the results 
are given for the optimal scale (and parameter $c$ in the case of
BSAF). The improvement of the BSAF with respect to the MF
ranges from $\sim 40\%$ (for white noise) to zero (for the largest
values of $\gamma)$.
We would also like to point out that the MH at the optimal scale
performs similarly to the standard MF. In addition, the MH has an
analytical expression which makes it very robust and easy to
implement. Therefore, it can be a useful filter in some practical
cases. 

We have also explored how the previous results change when varying 
$n^*_b$ and $R$. In particular, we have considered vaules of $n^*_b$
in the interval 0.01 - 0.05, $R=2$ and $R=3$ and values of $\gamma =
0, 0.5, 1$. The results are summarized in table~\ref{tab:tabla1_n}
for the BSAF and the standard MF (we present only those cases
where the BSAF improves at least a few per cent the standard MF).
The values of $\alpha$ and $c$ for the BSAF are found as the ones that
maximise $n^*$ in each case. 
\begin{table}
\begin{center}
\begin{tabular}{cccccccc}
\hline
$R$ &  $n^*_b$ & $\gamma$ & $\alpha$ & c &$n^*_{BSAF}$ & $n^*_{MF}$ &
$RD [\%]$ \\ 
\hline
\hline
2 & 0.01 & 0    & 0.4 & -0.69 & 0.0860 & 0.0824  & 4.4 \\
\hline
2 & 0.03 & 0   & 0.4  & -0.68 & 0.1493 & 0.1311 & 13.9 \\
  &      & 0.5 & 0.4 & -0.59 & 0.1512 & 0.1474 & 2.5 \\
\hline
2 & 0.05 & 0 & 0.4   & -0.70 & 0.1900 & 0.1575 & 20.6  \\
  &      & 0.5 & 0.4  & -0.59 & 0.1935 & 0.1783 & 9.0\\
\hline
3 & 0.01 & 0   & 0.3 & -0.86 & 0.0784 & 0.0658  & 19.1 \\
\hline
3 & 0.03 & 0  &  0.3 & -0.86  & 0.1282 & 0.1013  & 26.5 \\
& & 0.5  & 0.3  & -0.73 & 0.1242 & 0.1145 &  8.4 \\
\hline
3 & 0.05 & 0   & 0.3 & -0.86 & 0.1654 & 0.1186 & 39.4 \\
& & 0.5 & 0.3  & -0.75 & 0.1616 & 0.1352 & 19.5\\
& & 1   & 0.4 & -0.58  & 0.1582 & 0.1487 & 6.3 \\
\hline
\end{tabular}
\caption{Uniform  distribution. Number density of detections $n^*$ for
the standard MF ($\alpha=1$) and the  
BSAF with optimal values of c and $\alpha$. RD means relative
difference of number densities in percentage: $RD\equiv 100 (-1 +
n^*_{BSAF}/n^*_{MF})$.\label{tab:tabla1_n}} 
\end{center}
\end{table}
The relative performance of the BSAF with respect to the standard MF,
improves when $n^*_b$ increases. For instance, for $R=3$ and $\gamma=0$, the
improvement decreases from $\sim 40\%$ (for $n_b^*=0.05$) to $\sim 20\%$
(for $n_b^*=0.01$). On the other hand, as $R$ increases, the
difference between the detections found by both filters also
increases. In particular, for $n^*_b=0.05$ and $\gamma=0$, the
improvement goes up from $\sim 21\%$ (for
$R=2$) to $\sim 40\%$ (for $R=3$).

\subsubsection{Bright sources}

In order to test how the previous results are affected by the presence
of bright sources, we have also considered a uniform distribution that
contains a mixture of weak and bright sources with amplitudes in the interval
$A\in[0,5]\sigma_0$.
In particular, we have considered the reference example with values
$\gamma=0$, $n^*_b=0.05$ and $R=3$. For this case, we find that the optimal
values of the parameters for the BSAF are $c=-0.79$ and
$\alpha=0.3$. The behaviour is similar to the one 
found in the weak sources case, although the improvement of the BSAF
versus the MF is lower ($21\%$ as compared to the previous $40\%$).

It is interesting to note that for fixed values of the parameters
$n_b^*$, $\gamma$ and $R$, the optimal values of $\alpha$
and $c$ are very similar for both the weak and bright sources cases,
which is an indication of the robustness of the technique.

\subsection{Scale-free distribution}

\subsubsection{Weak sources}

For comparison purposes, we have repeated our analysis using 
a scale-free power-law distribution of sources with $\beta=0.5$ and
amplitudes in the interval $A\in[0.5,3]\sigma_0$,  
where $\sigma_0$ is the dispersion of the map
filtered with the standard MF. Therefore, we are considering the case
of weak sources since the corresponding upper limit for 
$\nu$ in the original (unfiltered) maps is below 3.


\begin{figure} 
\epsfxsize=75mm
\epsffile{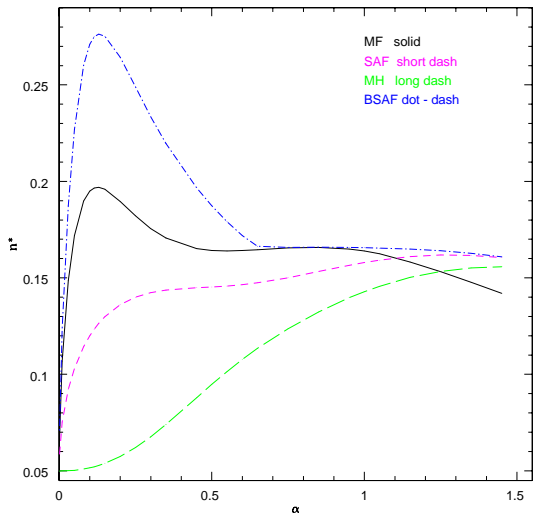}
\caption{Scale-free  distribution. The expected number density of detections $n^*$ as a function of the filter parameter $\alpha$ for 
$\gamma=0$  for the BSAF (using the optimal values of c), MF, SAF and MH filters. We consider the case $R=3$, $n^*_b=0.05$,  $\nu_i=0.5$, $\nu_f=3$ and $\beta=0.5$.}
\label{fig:numdet_alfa_pwr_g0}
\end{figure}
\begin{figure}
\epsfxsize=75mm
\epsffile{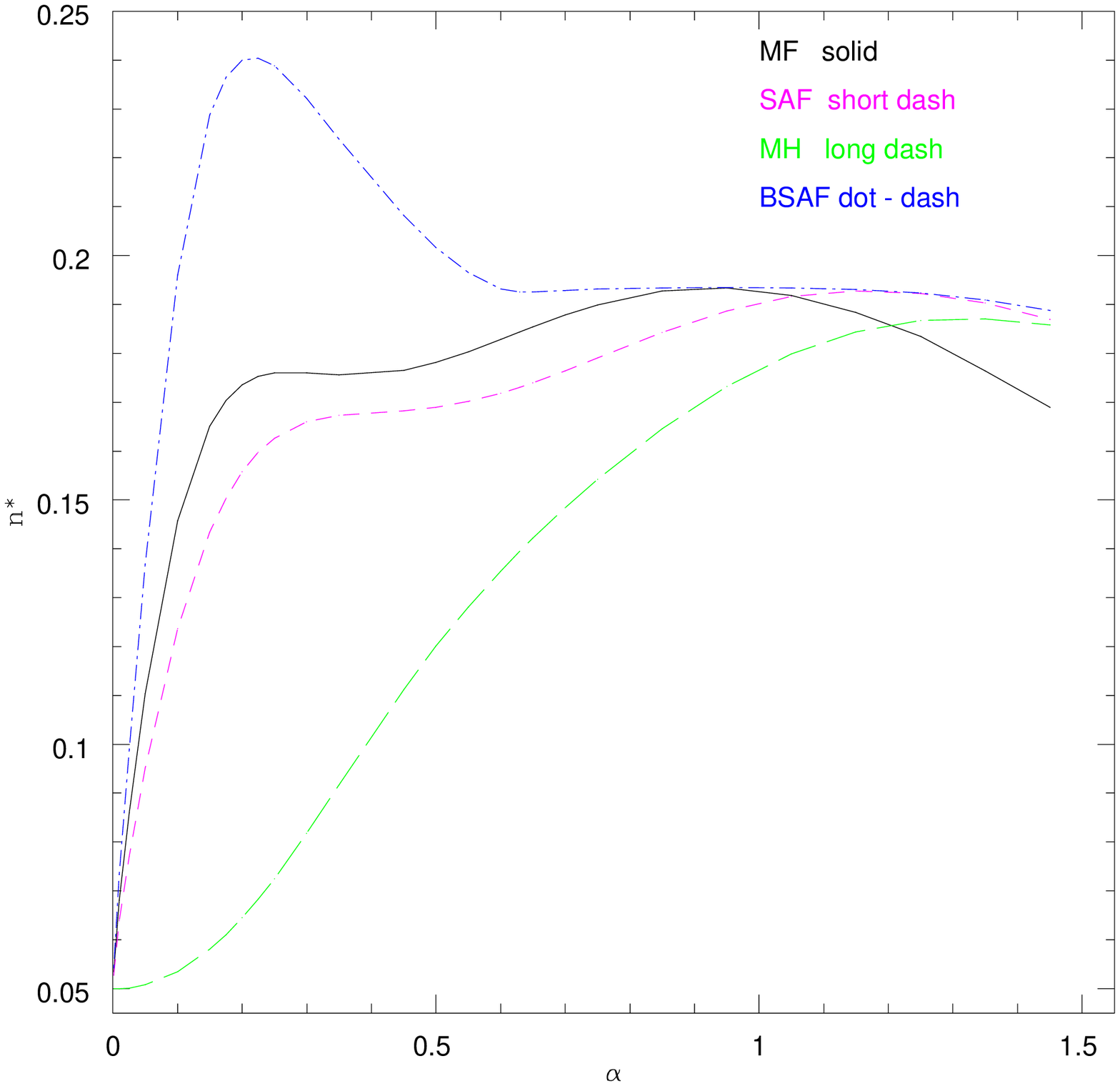}
\caption{Scale-free distribution. The expected number density of detections $n^*$ as a function of the filter parameter $\alpha$ for 
$\gamma=0.5$  for the BSAF (using the optimal values of c), MF, SAF and MH filters. We consider the case $R=3$, $n^*_b=0.05$,  $\nu_i=0.5$, $\nu_f=3$ and $\beta=0.5$.}
\label{fig:numdet_alfa_pwr_g05}
\end{figure}
\begin{figure}
\epsfxsize=75mm
\epsffile{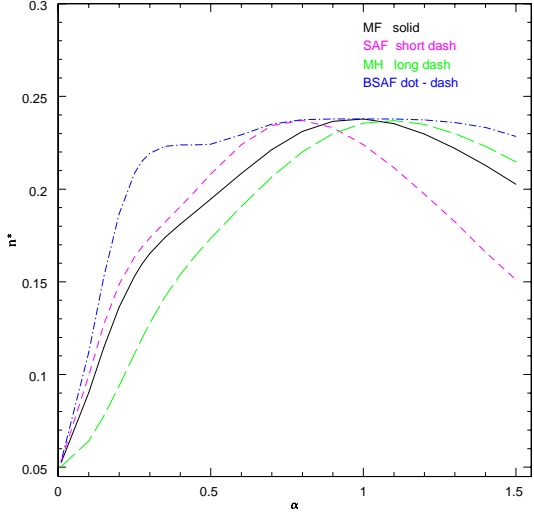}
\caption{Scale-free distribution. The expected number density of detections $n^*$ as a function of the filter parameter $\alpha$ for 
$\gamma=1.5$  for the BSAF (using the optimal values of c), MF, SAF and MH filters. We consider the case $R=3$, $n^*_b=0.05$,  $\nu_i=0.5$, $\nu_f=3$ and $\beta=0.5$.}
\label{fig:scale_g1p5}
\end{figure}
 
In figures \ref{fig:numdet_alfa_pwr_g0} and
\ref{fig:numdet_alfa_pwr_g05}, we plot $n^*$, the number  
density of detections, as a function of $\alpha$ for the cases
$\gamma=0$ and $\gamma=0.5$, assuming $R=3$ and $n^*_b=0.05$.  
In figure \ref{fig:numdet_alfa_pwr_g0}, $n^*$ is significantly higher
for the BSAF compared with the other filters  
at certain values of $\alpha$. In this case, the improvement of the
BSAF at $\alpha=0.3$ compared with
the standard MF is  $\simeq 42\%$. If  
we compare with the MF at $\alpha=0.3$, the improvement is  $\simeq
33\%$. In figure 
\ref{fig:numdet_alfa_pwr_g05}, with $\gamma=0.5$, an improvement of
$\simeq 20\%$ is obtained for the BSAF at $\alpha = 0.3$ with respect
to the standard MF.

As in the uniform distribution case, the BSAF gives very similar
results to the MF in the range $1 < \gamma \le 2$. 
In figure~\ref{fig:scale_g1p5} we show the results for $\gamma=1.5$,
$n_b^*=0.05$ and $R=3$. Again we see that the optimal BSAF defaults to the
standard MF.

\begin{figure} 
\epsfxsize=75mm
\epsffile{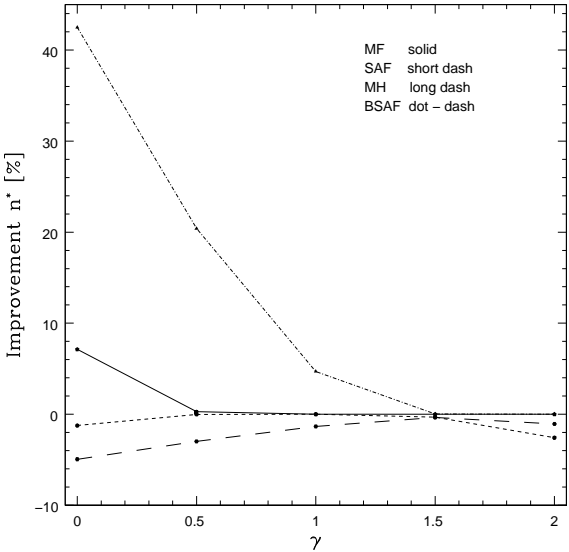}
\caption{Relative difference in the number of detections, 
with respect to the standard MF, as a function of the spectral
index $\gamma$. The results are shown for a scale-free distribution
($\beta=0.5$, $\nu_i=0.5$, $\nu_f=3$) for values of $n_b^*=0.05$ and $R=3$.}
\label{fig:exp_gamma}
\end{figure}
These results are summarized in Figure~\ref{fig:exp_gamma}, which
shows the relative difference in the number of detections, 
with respect to the standard MF, as a function of the spectral
index $\gamma$ for the different filters. At each point the optimal
scale (and parameter $c$ in the case of BSAF) has been used. We 
remark that for the interesting case of white noise more than a $40 \%$ of
detections is gained with respect to the standard MF. 

\begin{table}
\begin{center}
\begin{tabular}{cccccccc}
\hline
$R$ &  $n^*_b$ & $\gamma$ & $\alpha$ & c &$n^*_{BSAF}$ & $n^*_{MF}$ & $RD [\%]$ \\
\hline
\hline
2 &0.01 &0   & 0.4 & -0.66 & 0.1659 & 0.1590 & 4.3 \\
\hline
2 & 0.03 & 0   & 0.4 & -0.68 & 0.2376 & 0.2089 & 13.7\\
& & 0.5        & 0.4 & -0.56 & 0.2451 & 0.2432  & 7.8 \\
\hline
2 & 0.05 & 0   & 0.4 & -0.68 & 0.2772 & 0.2311 & 19.9\\
& & 0.5  & 0.4 & -0.57 & 0.2873 & 0.2705 & 6.2\\
\hline
3 & 0.01 &0   & 0.3 & -0.83 & 0.1336 & 0.1180 & 13.2\\
\hline
3 & 0.03 & 0    & 0.3 & -0.83 & 0.1975 & 0.1512 & 30.6\\
& &0.5          & 0.3 & -0.71 & 0.1937 & 0.1767 & 9.6\\
\hline
3 &0.05 & 0     & 0.3 & -0.81 & 0.2335 & 0.1639 & 42.5\\
& & 0.5         & 0.3 & -0.70 & 0.2321 & 0.1928 & 20.4\\
& & 1           & 0.3 & -0.62 & 0.2271 & 0.2169 & 4.7\\
\hline
\end{tabular}
\caption{Scale-free distribution. Number density of detections $n^*$ for the standard MF($\alpha$=1) and the BSAF with optimal values of c and $\alpha$. 
$RD\equiv 100 (-1 + n^*_{BSAF}/n^*_{MF})$.
\label{tab:tabla2_n}}
\end{center}
\end{table}
We have also explored how the results depend on the values of $R$ and
$n_b^*$.  In table~\ref{tab:tabla2_n}, we show the number density of
detections for the BSAF and for the standard MF  
($\alpha=1$) for $R=2$ and $R=3$, with $n^*_b$ ranging from 0.01 to
0.05, and for values of $\gamma=0,0.5,1$ (we only include the results
for those cases where the relative difference between the BSAF and
standard MF is at least a few per cent). We also give the optimal
values of $c$ and $\alpha$ where the BSAF performs better (taking 
into account the constraint $\alpha R \ga 1$).  
As for the previous case of the uniform distribution, the relative
performance of the BSAF improves when increasing $R$ and $n_b^*$.

It is also interesting to consider other values of the parameter
$\beta$. For instance $\beta \in [2.2,2.5]$ 
has an intrinsic interest for 
astronomy, because they describe the distribution of compact  
sources in the sky at microwave wavelengths. We have also explored 
the performance of the filters for these
values (for the reference case $\gamma=0$, $n_b^*=0.05$, $R=3$) and
the improvement of the BSAF (with optimal values of 
$\alpha=0.3$ and $c=-0.86)$ versus the standard MF is still  
significant and of the order $\simeq 35\%$.

\subsubsection{Bright sources}

To test the effect of the presence of bright sources on our results,
we have also considered a
scale-free distribution with $A\in[0.5,5] \sigma_0$ (i.e., a mixture
of weak and bright sources) for $\beta=0.5$. 
We find, for the reference case ($\gamma=0$, $n_b^*=0.05$, $R=3$), 
that the BSAF improves the standard MF around a $25\%$,
with optimal parameters $\alpha=0.3$ and $c=-0.76$.

We would like to point out that for for a given set of
$\gamma$, $R$ and $n^*_b$, this distribution of weak and bright
sources leads to very similar optimal parameters for the BSAF as the
scale-free distribution of weak sources.

In addition, we have also tested the performance of the filters 
for a scale-free distribution of bright sources with $A\in[3,5] \sigma_0$,
for the same case as before ($\gamma=0$, $n_b^*=0.05$, $R=3$). We
explore the  
parameter space of $(c,\alpha)$, looking for the best filter regarding
detection. We find that, for this distribution, the optimal parameters for
the BSAF are $c=0$ and $\alpha=1$, that is, the BSAF defaults to the
standard MF.

\subsection{On the robustness of the filters} \label{sec:robustness}

The filters considered here depend on a number of parameters
($\alpha$ in the case of SAF, MF and MH and $\alpha$ and $c$ 
in the case of BSAF) that must be determined in order to get
the maximum number of detections for a fixed number of spurious
detections. While for a given filter the region of acceptance 
is explicitly independent of the source distribution, the 
methodology presented here for the estimation of the optimal filter 
parameters depends on some assumed parameters of the source 
distribution (namely $\beta$, $\nu_i$ and $\nu_f$) and the noise
power spectrum ($\gamma$). A full study of the robustness of the
method for all the filters is out of the scope of this work. 
However, we have considered some interesting cases as tests of
the robustness of the method.

\subsubsection{Robustness with respect to the assumed source distribution}

In order to ascertain to what extent the uncertainties in
the $\beta$ parameter of the source distribution affects 
the determination of the optimal filter parameters, 
we repeated our calculations using wrong assumptions on
its value. An interesting case corresponds to assume
that the source distribution is uniform when, in reality, it is scale-free
and vice versa. 
In order to do this, we first construct the BSAF using the optimal
$(\alpha,c)$ values that were obtained for the uniform
distribution. We calculate then the number density of sources $n^*$ obtained 
from a map that contains sources that follow a scale-free distribution
with $\beta=0.5$. We find that the differences in the number
of detections when using the wrong filter with respect to
using the optimal one are very small (lower than $0.1 \%$). 
The same happens if the filter is constructed assuming an underlying
scale-free distribution and applied to a map with sources uniformly
distributed. 
This is not surprising, since tables~\ref{tab:tabla1_n} and~\ref{tab:tabla2_n}
show that both uniform and scale-free distributions
lead to similar values of the optimal $\alpha$ and $c$ 
parameters. 

Another source of uncertainty that appears in any real case is the
value of the limiting cut-offs of the source distribution, $\nu_i$ and
$\nu_f$. We have seen in the previous subsections that for a given set
of values $n_b^*$, $\gamma$ and $R$, different cut-offs for the 
same distribution lead to similar optimal $\alpha$ and $c$ parameters.
For instance, 
in our reference example ($\gamma=0$, $n_b^*=0.05$, $R=3$)
and a uniform distribution of sources with $\nu_i=0$ and $\nu_f=2$ the
optimal filter parameters are 
$\alpha=0.3$ and
$c=-0.86$, whereas if the upper cut-off value is
$\nu_f=5$, the
parameters take the values $\alpha=0.3$ and $c=-0.79$. 
Then, the shape of the optimal filter is only weakly dependent on the
value of the cut-offs. This suggests
that the methodology presented here is robust against uncertainties
in the prior knowledge of the cut-offs of the distribution.

In order to test this idea we proceeded in an analogous way to the
case of the $\beta$ parameter explained before: 
we apply wrong filters (that is, filters whose parameters have
been determined assuming wrong values of the cutoffs) to test cases
with real distributions of sources.
We tested several interesting cases: 
for uniform distributions, we studied the case
when the sources are assumed to be
weak but in reality some of them are bright (in our example, $\nu_f$ 
is assumed to be 2 but in reality its true value is $\nu_f=5$) and
the opposite situation. For the scale-free distribution, we 
studied the effect of mistaking the lower cut-off value (assuming
$\nu_i=0.5$ instead of its true value $\nu_i=3$ and vice versa). 
For all the cases, we plotted the curves $n_*$ versus $\alpha$. 
We observe that using a wrong filter changes the number of
detections of all the filters, but the qualitative behaviour
of the $n_*$ -- $\alpha$ curves does not change. The relative 
behaviour of the filters is basically the same, and therefore
the conclusions we obtained in the previous sections are still valid.

Thus, we conclude that the uncertainty in the knowledge of the
source distribution is, in general, not a critical issue in the cases
we have considered.

\subsubsection{Robustness with respect to the assumed power spectrum}

A more delicate issue is the one related to the assumption of the 
$\gamma$ parameter. If one assumes a value for $\gamma$ that is very
different from the true one, 
the shape of the filters changes dramatically
(except in the case of the MH whose shape is 
independent of $\gamma$) with respect to the optimal ones and this may
lead to wrong results. 
However, note that there are very well established techniques 
to estimate the power spectrum. Albeit in this academic case 
we consider power law-type backgrounds,
it is straightforward to apply the method to any kind of
power spectrum that can be present in the data.  

\begin{figure} 
\epsfxsize=84mm
\epsffile{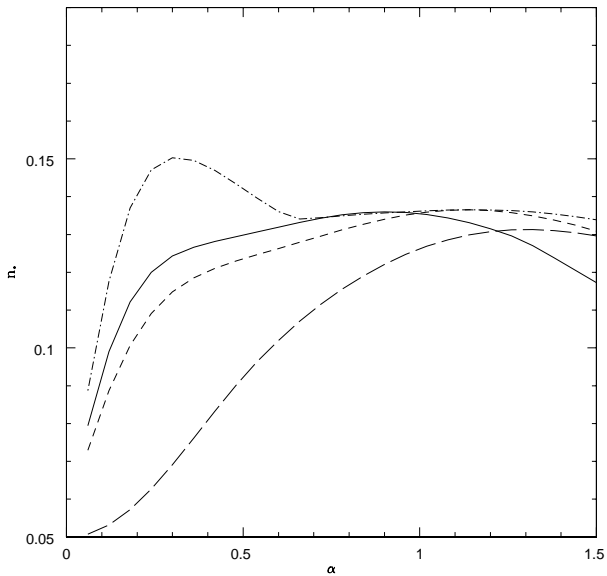}
\caption{Filter performances when the $\gamma$ parameter
of the background is poorly known. A uniform distribution is considered. 
The true value of the background power spectrum index is $\gamma_t=0.5$
whereas the filters have been constructed with a wrong $\gamma_w=0.6$.
Note how the figure compares with figure~\ref{fig:numdet_alfa_unif_g05}.}
\label{fig:gamma_falsa}
\end{figure}

As an example, we consider a case where the 
background corresponds to a true value $\gamma_t=0.5$
whereas the filters have been constructed
with a wrong $\gamma_w=0.6$, that is, a $20 \%$ error
in the determination of $\gamma$. The resulting $\alpha-n_*$
curves are given in figure~\ref{fig:gamma_falsa}.
The behaviour of the BSAF, the MF and the SAF is qualitatively
similar to the case where the noise power spectrum is perfectly 
known, but the performance of the three filters
is poorer. The MH curve is identical 
to the ideal case since the shape of the filter does not depend on $\gamma$. 
The BSAF still outperforms all the other filters, although the
improvement in the number of  
detections with respect to the MF slightly decreases.

\section{SIMULATIONS}

In order to see how our theoretical
framework works in a practical example,
we run a large set of simulations and study the
performance of the NP detector after filtering them
with the different filters considered in the previous sections.
We choose as a practical example the interesting
case of a Gaussian background characterised by a white 
noise power spectrum ($\gamma$=0) and sources whose
intensity distribution is uniform. We will focus on the 
detection of weak sources. 
For the sake of simplicity, we
give the 
results only for the BSAF and the MF since the other two filters 
(SAF and MH) perform worse 
in the considered case. 

\subsection{The simulations}

    The different simulations are performed as follows. The images 
    contain a number $N=4096$ pixels, which is sufficiently
    large so that 
    the addition 
    of a single source does not modify significantly the dispersion 
    of the images. The background is generated 
    as a random field with dispersion
    (before filtering)  
    $\sigma_0^{\rm unf} = 1$ (in arbitrary units).
 
    The sources that we have considered 
    for this example have a characteristic scale $R=3$ pixels. 
    Since we are interested in the detection of weak sources, 
    we add point sources with a uniform amplitude 
    distribution in the interval 
    $A \in [0,0.86]$ in the same arbitrary units
    of the background.
    The images filtered with the standard MF ($\alpha=1$)
    for this scale ($R=3$) 
    have dispersion $\sigma_{0} \simeq 0.43$.
    Thus, the sources are distributed in the interval
    $ \nu \in[0,2]$, where $\nu = A/\sigma_0$
    is the normalized amplitude of the sources with respect
    to the dispersion of the field filtered with the
    standard MF.

\subsection{Empirical NP criterion}

    For every maximum in a given image, it is possible to
    apply an empirical NP criterion to decide whether the maximum
    corresponds or not to a source. The quantities in equations (\ref{eq:r_*})
    and (\ref{eq:phi}) can be obtained from simulations in the
    following way:

    \subsubsection{Momenta, amplitude and curvature}

         The momenta
	 $\sigma_0$, $\sigma_1$  
	 and $\sigma_2$ (and, therefore, the
	 quantities $\rho$, $y_s$ needed  
	 to know the value of the linear detector $\varphi$)
	 can be straightforwardly calculated from the image.
	 For every
	 maximum in the image,  
	 it is possible as well to measure directly its amplitude $A$ and 
	 curvature $\kappa$. The normalized curvature is easily
	 obtained by Fourier
	 transforming the image, multiplying by $q^2$ and going back to
	 real space. This gives the value of $-\xi^{\prime \prime}$ at each
	 point and $\kappa$ is obtained dividing by $\sigma_2$.

    \subsubsection{Critical value $\varphi_*$}

         The critical value $\varphi_*$ that defines  
	 the acceptance region using equation (\ref{eq:r_*}) 
	 can be obtained as well directly from the simulations. 
	 For each considered filter, it is in principle possible to
	 calculate $\varphi_*$ semi-empirically, inverting
	 equation (\ref{eq:nb*}) (with the empirically 
	 obtained values of $\rho$ and $y_s$) just as we did in the previous 
	 sections, and hence to proceed with the NP decision rule. 
	 Instead, since we are dealing with simulations, we will follow a fully empirical approach.

	 The argument goes as follows. We fix the  number density 
	 of spurious detections, i.e., the number of maxima of the 
	 background that are misidentified as ``sources" by our 
	 detection criterion. Then we simulate a set of images 
	 containing only background and filter them with 
	 the filter under study. We focus on the background maxima 
	 and try to determine the value of $\varphi_*$ that 
	 makes the NP rule to produce the specified number of 
	 spurious detections. For example let us consider that
	 we perform $N_{tot}=50000$ 
	 noise realisations and focus on what happens in a 
	 certain pixel (we choose the central pixel of the 
	 simulation in order to avoid border effects). For every 
	 realisation, we check if there is a maximum at this 
	 position or not. If a maximum is present, the value of 
	 $\varphi$ is calculated. All the values of $\varphi$ 
	 obtained in that way are sorted into descending numbers 
	 (large to small). The value of $\varphi_*$ is then given 
	 by the $\varphi$ corresponding to the r-th element $(r=n^*_b
	 N_{tot})$ of the sorted list (that is, $\varphi_*$ is the  
	 value of $\varphi$ so that there are $n^*_b N_{tot}$ 
	 background maxima with $\varphi \geq \varphi_*$). 
	 For this example, we considered $n^*_b=0.05$ and therefore 
	 $n^*_b N_{tot}=2500$.

    \subsubsection{Number of detections}

         Once the value of $\varphi_*$ has been empirically determined
	 we add a source 
	 with a Gaussian profile of dispersion $R = 3$ pixels at 
	 the central position (pixel $N/2+1$) of the unfiltered 
	 background image and then 
	 we filter it with the considered filter.
	 We proceed to apply the detector to any maxima 
	 located at the pixel we are 
	 considering. Finally, $n^*$, for each 
	 of the filters, will be the number of sources with an estimated  
	 $\varphi \geq \varphi_*$  divided by 
	 the total number of realizations  $N_{tot}$.

\subsection{Results}

    We have performed numerical simulations for our 
    reference case ($n^*_b=0.05$,  $R=3$ and $\gamma=0$), assuming a
    uniform distribution with  $\nu_c=2$. We have compared the
    performance of the filters and the empirical NP
    detector in the simulated images with the theoretical predictions
    as a function of the parameter $\alpha$. 
    For each $\alpha$ value and each filter,
    we have 
    done five sets of simulations of the background, 
    each one with a number
    of realizations 
    large enough 
    to have 5000 of them 
    containing a maximum in the 
    central pixel\footnote{The total number of realizations in
    each set of simulations needed to obtain 5000 of them 
    with a maximum
    in the central pixel 
    was 
    $\simeq 50000-60000$.}. 
    We used these simulations
    to obtain $\varphi_*$ as explained before. 

\begin{figure} 
    \epsfxsize=85mm 
    \epsffile{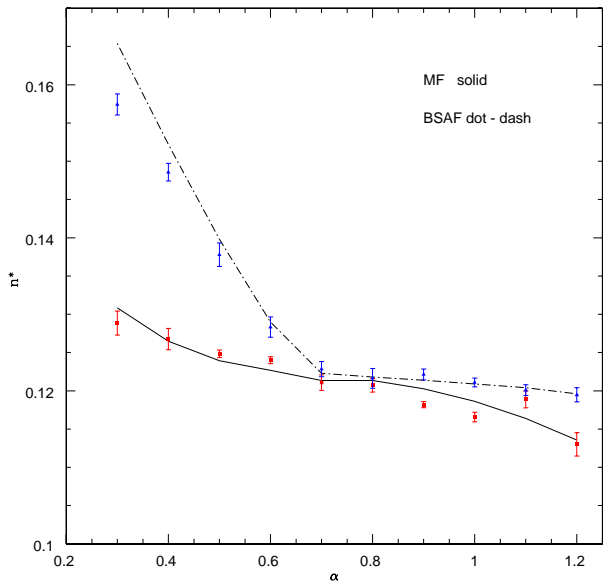} 
    \caption{Number density of detections from theory and simulations
      versus the filter parameter $\alpha$ for 
      the BSAF and MF. The solid and dot-dash lines 
      represent the theoretical number density $n^*$ and the 
      squares and triangles are the results from the 
      simulations for $\nu_c=2$, $n^*_b=0.05$, 
      $R=3$ and $\gamma=0$, for the MF and the BSAF respectively.} 
    \label{fig:numdet_sims_todo} 
    \end{figure}

    In the figure~\ref{fig:numdet_sims_todo}  
    we present the results from the simulations for this case and the
    comparison  
    with the theoretical calculations.  
    The lines in this plot show the theoretical results for each filter, 
    the triangles the result from the simulations 
    for the BSAF and the squares the results for the MF. We concentrate 
    on the BSAF, which corresponds to the dot-dash line. As we mentioned 
    in previous sections, the 
    BSAF significantly improves 
    the standard MF for $\gamma=0$.  The simulations 
    follow the theoretical results well. In the region where 
    $\alpha \simeq 0.3$, there is a small deviation from theory 
    which we believe is related to the fact that we are close to 
    the scale of the pixel, but still, significantly close to the 
    expected theoretical value.

\subsection{The estimation of the amplitude of the source}

\begin{figure} 
\includegraphics[width=6cm,angle=-90]{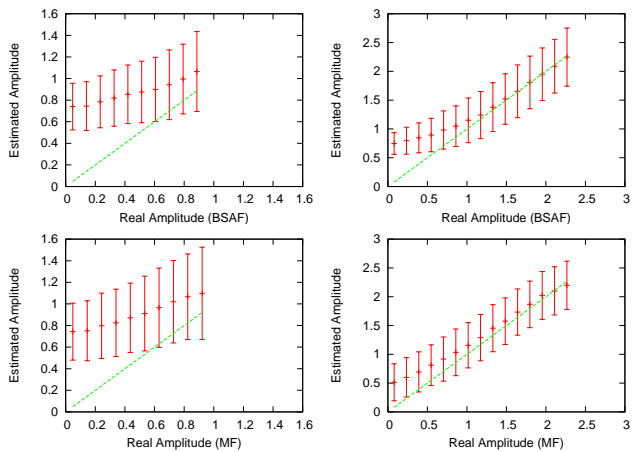}
\caption{Real versus estimated amplitudes of
  simulated sources for the BSAF (top panels) and the standard MF
  (bottom panels). Two cases have been considered: left panels show
  the results for sources uniformly distributed in the interval
  $A\in[0,2]\sigma_0$ whereas right panels show the results for
  sources distributed in $A\in[0,5]\sigma_0$. For all four
  cases, the parameters of the simulations are
  $n^*_b=0.05$, $\gamma=0$ and $R=3$ pixels.}
\label{fig:estim_ampli1}
\end{figure}

We can estimate the amplitude of a source using the unbiased and
maximum efficient estimator from equation (\ref{eq:estim_amplitude})
and then compare it with the amplitude that we have randomly
generated. In figure \ref{fig:estim_ampli1}, we plot the real
amplitude versus the estimated one for the BSAF (top panels) and the
MF (bottom panels). 

Two amplitude regimes have been explored. 
In the left panels, we have simulated a uniform distribution of
weak
sources, $A\in[0,2]\sigma_0$. The parameters used for these simulations
are $n^*_b=0.05$, $\gamma=0$ and $R=3$. 
The optimal filter parameters have been chosen at each case.
The points and the error bars are
calculated as the average and the dispersion of the detected sources
that fall in each of the amplitude bins from a total of  $\simeq$
10000 detected sources. We find a similar positive bias in the
determination of the amplitude for the BSAF ($\alpha = 0.3$,
$c=-0.86$) and MF ($\alpha = 0.3$). However, the error bars
corresponding to the BSAF are slightly smaller than those of the MF. 

In the right panels, we give the results for a uniform distribution of
sources with $A\in[0,5]\sigma_{0}$. 
The simulation parameters are the same as before ($n^*_b=0.05$,
$\gamma=0$ and $R=3$). As before, the points and
the error bars are calculated as the average and the dispersion of the
detected sources that fall in each of the amplitude bins from a total
of  $\simeq$ 20000 detected sources. We find that the BSAF
($\alpha=0.3$, $c=-0.79$) is unbiased for bright sources 
whereas the estimation of the amplitude in the
case of the standard MF ($\alpha=1$)
shows some bias even for bright sources.
Therefore,
the BSAF with $\alpha=0.3$ outperforms the standard MF
in both the detection and estimation for this distribution.

The fact that sources with small amplitudes are significantly affected
by a positive bias can be explained taking into account 
that these sources are more easily detected if they lie over a
positive contribution of the background. This contributes 
systematically to the overestimation of the amplitude. We would like
to point out that this 
estimator produces appreciably better results than a naive estimation
using directly the measured values at the maxima.

\section{CONCLUSIONS}
Nowadays, the detection of compact sources on a background is a
relevant problem in many fields of science. A   
number of detection techniques use linear filters and
thresholding-based detectors.  
Our approach to the problem of detector design is different. We use a
Neyman-Pearson  
rule that takes into account {\it a priori} information of the
distribution of sources and the number density  
of maxima to define the region of acceptance.

In our case, we take advantage not only of the
amplification but also of      
the spatial information: the curvature of the background is different 
from that of the sources, and we use this to improve our detection
rule. 

The background is modelled by
a homogeneous and isotropic Gaussian random
field, characterized by a scale-free power spectrum $P(q) \propto
q^{-\gamma}$, $\gamma \geq 0$. 

We design a new filter that we call BSAF in such a way that the use of
our improved detection rule based on amplification and  
curvature on the filtered field will increase the number of detections
for a fixed number of spurious 
sources. We generalize the functional form of this filter, as well as
other standard filters, and introduce another degree of freedom,  $\alpha$,  
that allows us to filter at any scale, including that of the source
$R$. We have shown the benefits of  
filtering at scales smaller than $R$, which significantly improves the
number of detections. 

As an example, we have considered two different distributions  
of sources. A uniform distribution in the interval $ \nu \in[0,2]$ 
and a scale-free power law distribution in the interval $ \nu
\in[0.5,3]$ (where the threshold $\nu$ corresponds to the field
filtered with the standard MF), i.e. we are considering weak sources.
The BSAF 
has proven to be significantly better than the standard MF, the  SAF
and MH wavelet in certain cases. In particular, we have considered a
reference case with parameters $\gamma=0$, $n_b^*=0.05$ and $R=3$,
where the improvement in the
number of detections of the BSAF at $\alpha=0.3$ with respect to the
standard MF is $\simeq 40\%$. We have also tested the performance
of the filters for a mixture of weak, intermediate and bright
sources. For a uniform distribution with   
$\nu\in[0,5]$ and for a scale-free distribution with 
$\nu\in[0.5,5]$, the BSAF also improves the MF. However, 
for a scale-free distribution with $\nu\in[3,5]$, i.e.,
dominated by bright sources, we find that the optimal 
BSAF defaults to the standard MF, which gives the maximum number of
detections in this case. 

We find that the BSAF gives in any case the best performance
among the considered filters. Indeed, the SAF and the MF are particular cases 
of the BSAF and the strategy  we follow, i.e. 
maximization of the detections, guarantees that the parameters of the
BSAF will default to the best possible of these filters in each case. 
In addition, we also find that the BSAF performs at least as well as the MH in all the
considered cases. Therefore, the number
density of detections obtained with the BSAF will be at least equal
to the best of the other three filters, and in certain cases superior.
However, in some other cases, the gain is small and it is justified
to use an analitically simpler filter. 
Our results suggest that for power law spectra, from the practical point of view, 
one could use the BSAF when $0 \la \gamma \la 1$ since, in this range, clearly
improves the number of detections with respect to the other filters. However,
for  $\gamma \ga 1.0$ the usage of the MH is justified due to its robustness 
(since it has an analytical form) and it gives approximately the same number of
detections obtained either with the BSAF or MF.

For all the studied cases of source distributions (except for the one
dominated by bright sources) and fixing the values of $\gamma$,
$n_b^*$ and $R$, we find that the optimal
parameters of the BSAF are only weakly dependent on the 
distribution of the sources.
We have done some simple tests in order to study the robustness
of the method when the knowledge about the source {\it pdf} or the
background spectral index is not perfect. We find that 
the values of the optimal filter parameters vary slightly 
when we assume
that the source distribution is uniform when, in reality, it is scale-free
and vice versa. The uncertainties in the cut-off values of the
source {\it pdf} affect the number of detections, but in a
similar way for all the filters, and therefore the relative
behaviour of the filters do not change. Errors in the 
estimation of the spectral index $\gamma$ reduces the 
efectiveness of the BSAF, but it still outperforms the other
filters. All of this indicates that our detection scheme
is robust against uncertainties in the knowledge
of the distribution of the sources and spectral index.

To test the validity of our results in a practical example, 
we have tested our ideas with simulations for the uniform
distribution (using our reference case $n_b^*=0.05$, $R=3$,
$\gamma=0$) and find that the results 
follow approximately the expected theoretical values.  

Regarding source estimation, we propose a
linear estimator which is unbiased and of maximum efficiency, that we have
also tested with simulations.  

The ideas presented in this paper can be generalized: application to other
profiles  (e.g. multiquadrics, exponential) and non-Gaussian
backgrounds is physically and 
astronomically interesting. The extension to include several images
(multi-frequency) is relevant. The generalization to two-dimensional
data sets (flat maps and the 
sphere) and nD images is also very interesting. Finally the
application of our method to 
other fields is without any doubt. We are currently doing research in
some of these topics.

\section*{Acknowledgements}

The authors thank Enrique Mart\'\i nez-Gonz\'alez and Patricio Vielva
for useful discussions. 
MLC thanks the Ministerio de Ciencia y Tecnolog\'\i a (MCYT) for a
predoctoral FPI fellowship. 
RBB thanks the MCYT and the Universidad de Cantabria for a Ram\'on y
Cajal contract. 
DH acknowledges support from the European Community's Human Potential
Programme 
under contract HPRN-CT-2000-00124, CMBNET. We acknowledge partial
support from  
the Spanish MCYT project ESP2002-04141-C03-01 and from the EU Research
Training Network `Cosmic Microwave Background 
in Europe for Theory and Data Analysis'.

\appendix

\section{}

The ratio $L(\nu ,\kappa |\nu_s) \equiv  n(\nu ,\kappa |\nu_s)/n_b(\nu ,\kappa )$ can be explicitly written as
\begin{equation}
L(\nu ,\kappa |\nu_s) = e^{\varphi \nu_s - \frac{1}{2}(\mu + y^2_s)\nu_s^2} ,
\end{equation}
and taking into account the NP criterion for detection, we find
\begin{equation}
\L(\nu ,\kappa )\equiv  \int_0^{\infty} d\nu_s\, p(\nu_s)  L(\nu ,\kappa |\nu_s) \geq L_*,
\end{equation}
where $L_*$ is a constant. By differentiating the previous equation with respect
to $\varphi$
\begin{equation}
\frac{\partial L}{\partial \varphi } = 
\int_0^{\infty} d\nu_s\, p(\nu_s)\nu_s e^{\varphi \hat{\nu}_s - \frac{1}{2}(\mu +
y^2_s)\hat{\nu}_s^2}\geq 0.
\end{equation}
Therefore, $\L(\nu,\kappa)\geq L_*$ is equivalent to $\varphi \geq \varphi_*$, where
$\varphi_*$ is a  constant, i.e. $\varphi (\nu ,\kappa )$ given by equation (\ref{eq:phi}) is a sufficient linear detector.

\section{}

Let us assume a linear estimator combination of the normalized amplitude $\nu$ and normalized 
curvature $\kappa$ with the constraint
\begin{equation}
{\hat{\nu}}_s = A\nu + B\kappa .
\end{equation}
If the estimator is unbiased, i.e. $\langle {\hat{\nu}}_s\rangle = \nu_s$,
taking into account that 
$\langle \nu\rangle = \nu_s$ and $\langle \kappa\rangle = \nu_sy_s$, we obtain
the constraint
\begin{equation} \label{eq:constraint_appc}
A + By_s = 1.
\end{equation}
On the other hand, the variance is given by
\begin{equation}
\sigma^2_{{\hat{\nu}}_s} = A^2 + B^2 +2\rho AB, 
\end{equation}
where we have taken into account that $\sigma^2_{\nu} = \sigma^2_{\kappa} = 1, \ 
 \langle \nu \kappa\rangle = \rho + y_s\nu^2_s$. 
By minimizing the previous expression with the constraint
(\ref{eq:constraint_appc}), 
one obtains
\begin{equation}
A = \frac{1}{y^2_s + \mu }\frac{1 - \rho y_s}{1 - \rho^2},\ \ \ 
B = \frac{1}{y^2_s + \mu }\frac{y_s - \rho}{1 - \rho^2},
\end{equation}
Therefore, one obtains:
\begin{equation}
{\hat{\nu}}_s = \frac{\varphi}{y^2_s + \mu },
\end{equation}
\begin{equation}
\sigma^2_{{\hat{\nu}}_s} = \frac{1}{y^2_s + \mu}.
\end{equation}

\end{document}